\documentclass{IOS-Book-Article}

\usepackage{mathptmx}
\usepackage{fancyvrb}
\usepackage{graphicx}
\usepackage{csquotes}
\usepackage{comment}
\usepackage{soul}\setuldepth{article}
\def\hb{\hbox to 11.5 cm{}}

\begin{document}

\pagestyle{headings}
\def\thepage{}
\begin{frontmatter}              % The preamble begins here.

%\pretitle{Pretitle}
\title{Toward FAIR Semantic Publishing of Research Dataset Metadata in the \\ Open Research Knowledge Graph}

%\subtitle{Toward FAIR Semantic Publishing of Research Dataset Metadata}

\author[A]{\fnms{Raia} \snm{Abu Ahmad}\orcid{0009-0004-8720-0116}%
\thanks{Corresponding Author: Raia Abu Ahmad, raia.abu\_ahmad@dfki.de.}},
\author[B]{\fnms{Jennifer} \snm{D'Souza}\orcid{0000-0002-6616-9509}},
\author[C]{\fnms{Matthäus} \snm{Zloch}\orcid{0009-0007-1692-7053}},
\author[C]{\fnms{Wolfgang} \snm{Otto}\orcid{0000-0002-9530-3631}},
\author[A]{\fnms{Georg} \snm{Rehm}\orcid{0000-0002-7800-1893}},
\author[B]{\fnms{Allard} \snm{Oelen}\orcid{0000-0001-9924-9153}},
\author[C]{\fnms{Stefan} \snm{Dietze}\orcid{0009-0001-4364-9243}}
\author[B]{\fnms{Sören} \snm{Auer}\orcid{0000-0002-0698-2864}}

%\runningauthor{Abu Ahmad et al.}
\address[A]{DFKI GmbH -- Deutsches Fourschungszentrum für Künstliche Intelligenz, Berlin, Germany}
\address[B]{TIB -- Leibniz Information Centre for Science and Technology, Hannover, Germany}
\address[C]{GESIS -- Leibniz Institute for the Social Sciences, Köln, Germany}

\begin{comment}
\author[A]{\fnms{First} \snm{Author}\orcid{....-....-....-....}%
\thanks{Corresponding Author: Author Name, contact details.}},
\author[B]{\fnms{Second} \snm{Author}\orcid{....-....-....-....}}
and
\author[B]{\fnms{Third} \snm{Author}\orcid{....-....-....-....}}

\runningauthor{B.P. Manager et al.}
\address[A]{Short Affiliation of First Author}
\address[B]{Short Affiliation of Second Author and Third Author}
\end{comment}

\begin{abstract}
Search engines these days can serve datasets as search results. Datasets get picked up by search technologies based on structured descriptions on their official web pages, informed by metadata ontologies such as the \textit{Dataset} content type of \textit{schema.org}. Despite this promotion of the content type \textit{dataset} as a first-class citizen of search results, a vast proportion of datasets, particularly research datasets, still need to be made discoverable and, therefore, largely remain unused. This is due to the sheer volume of datasets released every day and the inability of metadata to reflect a dataset's content and context accurately. This work seeks to improve this situation for a specific class of datasets, namely research datasets, which are the result of research endeavors and are accompanied by a scholarly publication. We propose the \textit{ORKG-Dataset} content type, a specialized branch of the Open Research Knowledge Graoh (ORKG) platform, which provides descriptive information and a semantic model for research datasets, integrating them with their accompanying scholarly publications. This work aims to establish a standardized framework for recording and reporting research datasets within the \textit{ORKG-Dataset} content type. This, in turn, increases research dataset transparency on the web for their improved discoverability and applied use. In this paper, we present a proposal -- the minimum FAIR, comparable, semantic description of research datasets in terms of salient properties of their supporting publication. We design a specific application of the \textit{ORKG-Dataset} semantic model based on 40 diverse research datasets on scientific information extraction.
%This work examines approaches to semantic representations of research dataset metadata. We argue that the current standards lack descriptions of the content and context of datasets, and thus a vast proportion of research datasets still need to be made discoverable and largely remain unused. Therefore, we propose the \textit{ORKG-Dataset}, a content type that facilitates detailed FAIR semantic descriptions of research datasets, supported by semantic scholarly publishing technologies. Using this content type, we aim to increase research dataset transparency on the web for their improved discoverability and applied use. We exemplify our design with a use-case of 40 different research datasets on scientific information extraction.

\end{abstract}

\end{frontmatter}
\markboth{July 2023\hb}{July 2023\hb}

\section{Introduction}

Scientific research has long been conducted on datasets that measure and model aspects of the world. This phenomenon is becoming more acute given the vast amounts of datasets \cite{cafarella2011structured, manyika2014open} released in the present age of data science \cite{verhulst2016open}. Datasets are as diverse as science is \cite{gregory2018understanding, mayer2013big}. For instance, datasets in medicine can reflect patient disease histories, datasets in earth science can reflect geological or climatic features of the world, while datasets in artificial intelligence model phenomena as machine learning objectives for a computer. 
The large volume of datasets released on the Web opens up new avenues for their search and discovery. Search engines now offer dedicated search platforms such as Google Dataset Search \cite{brickley2019google} for discovering datasets from various public repositories such as OpenAIRE \cite{manghi2019openaire} and Zenodo\footnote{\url{https://zenodo.org}}. To make datasets discoverable, dataset publishers must offer the dataset metadata information per certain prescribed formats. For example, Google advocates the \textit{Dataset}\footnote{\url{https://schema.org/Dataset}} type. However, many research datasets\footnote{Note in this work we draw a distinction between the generic concept of a dataset on the Web and research datasets, in particular. Research datasets are those that are outcomes of a research endeavor and are thereby accompanied by a scholarly publication.} do not come with a sufficient amount of structured metadata that fully describe their usage potentials. Therefore, discovery mechanisms used by search engines nowadays cannot detect and expose them to users. We have observed that metadata such as title, publisher, etc., contingent on their accuracy and maintenance \cite{chapman2020dataset}, are insufficient to describe a research dataset's full context and content. Therefore, such limited metadata proves fairly uninformative to guarantee that relevant research datasets will be discovered on the Web.

Studies show that, in academia, the predominant search pattern for research datasets is either a serendipitous event of finding a dataset when reading scholarly publications or actively searching for datasets 
in publications \cite{gregory2018understanding}. This emphasizes that the content of scholarly publications, which gives insights into datasets, 
is necessary for dataset discovery \cite{gregory2018understanding}. Prior work highlights three criteria based on which users select datasets: \textit{relevance}, \textit{usability}, and \textit{quality} \cite{koesten2020everything}. These factors, resulting from human data interaction studies \cite{piwowar2010public,boukhelifa2017data,gregory2019searching, kern2015there, thoegersen2021researcher, koesten2017trials}, necessitate the supply of content and context information of a dataset (e.g. domain(s) the dataset covers, source(s) it was gathered from, and metrics to evaluate it) to support informed decisions of its use for a task. We observe that such information is found in the scholarly publications that describe research datasets \cite{gregory2018understanding}. 

Therefore, the way forward toward improved research dataset discovery is to complement its metadata with a structured representation of the contributions of its accompanying scholarly article. To adhere to current standards of semantic descriptions, the representation of these contributions should be findable, accessible, interoperable, and reusable, i.e. it should adhere to the 
FAIR principles \cite{wilkinson2016fair}. To this end, semantic publishing models \cite{shotton2009semantic} of scholarly contributions such as the Open Research Knowledge Graph (ORKG) can be directly leveraged \cite{auer2020improving}. The ORKG publishing model presents a next-generation skimming device of scholarly contributions, that permits viewing their semantic representations in a similar way to comparisons of products on e-commerce websites \cite{oelen2020}. 
Thus, to model research datasets, we utilize the ORKG content type, which is a typed resource with a class from a predefined set of classes. 
%The ORKG content type facilitates class findability and the design of specialized user interfaces. Consequently, we propose a specialized branch of the ORKG: the ORKG-Dataset content type.

% The ORKG-Dataset seeks to offer a framework that supports the comprehensive semantic representation of research datasets. 

In this paper, we present the ORKG-Dataset content type -- a specialized branch of the ORKG. The design of ORKG-Dataset was driven by three main research questions (RQs). \textbf{RQ1:} How to present structured research dataset descriptions within the semantic web scholarly publishing model as knowledge graphs (KGs)?; \textbf{RQ2:} Which salient features can be extracted from scholarly article descriptions that serve the dataset selection criteria of \textit{relevance}, \textit{usability}, and \textit{quality}?; and \textbf{RQ3:} How can such a representation benefit others in terms of creating customizable snapshots of specific information? 

The rest of this paper is structured as follows: Section \ref{design-principles} presents the design principles of ORKG-Dataset and how they were met within the ORKG, addressing RQ1. Section \ref{application} demonstrates an application example of ORKG-Dataset on the scholarly publications of 40 research datasets used for scientific information extraction (IE), addressing RQ2 and RQ3. Finally, Section \ref{conclusion} concludes our paper.

%The characterization of datasets across the research fields in which they are created understandably varies in terms of their salient aspects which the ORKG-Dataset content type is designed to model. This work seeks to expand our understanding of how people can interact with and communicate about datasets. It can potentially impact the design of FAIR research-community-oriented representation standards of dataset descriptions in the semantic web. 

%Automated language technologies can also be geared to accommodate enriched templated patterns for dataset retrieval.

%This work is part of the project NFDI for Data Science (DS) and Artificial Intelligence (AI)--NFDI4DS\footnote{\url{https://www.nfdi4datascience.de/}}--, which is one of the initiatives in the German national research data infrastructure or \textit{Nationale Forschungsdateninfrastruktur} (NFDI) project, aiming to support all steps of the research data life cycle including collecting, creating, comparing, processing, analyzing, publishing, archiving, and reusing resources in DS and AI.   

\section{Design Principles of ORKG-Dataset}
\label{design-principles}

Although previous work has been conducted to describe the content and context of datasets, it is not fine-grained enough in terms of the properties it offers. For example, the Data Source Description Vocabulary~\footnote{\url{https://dqm.faw.jku.at/ontologies/dsd/4.0.0/index.html}} and the Data Catalog Vocabulary (DCAT)~\footnote{\url{https://www.w3.org/TR/vocab-dcat-3/}} hav no possibility to describe models trained on the dataset and their evaluation scores. Additional resources for describing datasets such as releasing datasheets \cite{gebru2021datasheets} are not modeled using semantic web technologies that enhance the usability and discoverability of datasets. 

%Keeping that in mind, we motivate the ORKG-Dataset design with the help of the following scenario. 

With this outlook, we outline the principles of ORKG-Dataset design using the following scenario. Imagine a researcher looking for research datasets in a particular field. Her search can be characterized based on two activities elicited in prior work \cite{koesten2017trials}, (1) linking (i.e. \enquote{finding commonalities and differences between two or more datasets} \cite{koesten2017trials}) and (2) time series analysis (i.e. ordering datasets on a timeline). %In this context, she might like to compare the research datasets addressing the same problem, or to have a performative overview of the reported performance benchmarks. 
In the present status quo of scholarly communication, a large share of the information discussed above is already available, but it is hidden within the unstructured text of scholarly articles accompanying the research datasets. However, searching for a dataset by examining unstructured text descriptions involves significant cognitive tie-ups which boils down to finding a needle in a haystack. The researcher would need to sift through millions of results from academic search engines, identify those that actually contribute a dataset, and tediously search through the articles for key information before finding a suitable dataset. 

The ORKG-Dataset content type introduced in this paper is one step in a long-term research agenda of the ORKG to bring about a paradigm shift from document-based to structured knowledge-based scholarly communication \cite{auer2020improving}. Specifically, the salient aspects of research are encoded as structured property-value pairs within the ORKG. Given the salient structured format applied on scholarly communication, the two search characteristics, i.\,e., linking and time series analysis, are directly supported in the interaction mechanisms of the ORKG front-end interface. Several structured papers with similar properties can be combined and placed next to each other within a comparison view \cite{oelen2020}. The ORKG platform combines semantic web technologies with front-end design components and back-end storage and query systems \cite{jaradeh2019open}. It utilizes Resource Description Framework (RDF) as its default graph data representation language\footnote{\url{https://www.w3.org/RDF/}}, connecting ontologies through subject-predicate-object triples. 
The front-end interfaces are built with ReactJS, fetching data from back-end APIs, while the Neo4J storage software\footnote{\url{https://neo4j.com}} enables effective data querying using SPARQL\footnote{\url{https://www.w3.org/TR/rdf-sparql-query/}}. 

Our proposed design requires each paper contribution to be typed as both the default \url{https://orkg.org/class/Contribution} class and the \url{https://orkg.org/class/Dataset} class. This is critical to separate other kinds of research contributions from research dataset contributions in the ORKG. %Furthermore, within the ORKG, several structured papers with similar properties can be combined and placed next to each other to form a comparison view \cite{oelen2020}. 
We identified the following design requirements in order to generate comparable and wholesome dataset structured representations as the ORKG-Dataset content type.

\begin{itemize}

    \item \textbf{Standardized Nomenclature:} We established a standard nomenclature for research datasets, starting with a controlled vocabulary that can later evolve into an ontology. This is achieved by reusing concepts from existing metadata ontologies such as \url{https://schema.org/Dataset}. We then added predicates specific to research datasets in the ORKG web namespace to enhance the vocabulary. To establish further equivalences between predicates in different ontologies, the RDF \textit{same-as} relation was utilized.

    \item \textbf{Use of Templates}: to maintain consistent formatting when recording new research datasets, it was essential to define a form-based template comprising a set of predetermined relevant predicates. The ORKG facilitates this requirement by implementing a template system that consists of recurring subgraph property patterns~\footnote{For more information about templates in the ORKG: \url{https://orkg.org/about/19/Templates}}. These templates enable the specification of commonly applicable properties across multiple research contributions within a KG. 

    \item \textbf{FAIR Standards Compliance}: the third and final requirement is that the ORKG-Dataset model should be compliant as much as possible with the FAIR (Findable, Accessible, Interoperable, and Reusable) guiding principles laid out for scientific information~\cite{wilkinson2016fair}. 
    Addressing \textit{findability}, the ORKG has a system to assign digital object identifiers (DOIs) to aggregate component parts of its graph data, making them available in global scholarly infrastructures such as DataCite and Crossref \cite{haris2022persistent}. ORKG resources are also findable via regular search engines due to being published on the web. In terms of \textit{accessibility}, all ORKG components are accessible by HTTP protocol via REST or its user interface. Additionally, the actual graph data is separated from metadata making it possible to access the latter without the former \cite{oelen2020}. The ORKG satisfies \textit{interoperability} by using RDF, the recommended format of the W3C for representing knowledge on the web in a machine-accessible format. To ensure \textit{reusability}, provenance metadata information is created automatically when publishing structured contributions in the ORKG, and the graph data is published as CC BY-SA.\footnote{\url{https://creativecommons.org/licenses/by-sa/2.0/}}

\end{itemize}

\section{The ORKG-Dataset Application}
\label{application}

In this section, we demonstrate a use-case application of the ORKG-Dataset content type on research datasets in the field of natural language processing (NLP) on the research problem of scientific information extraction.

\subsection{Datasets Curation}

As a first step, we manually curated a collection of research datasets relevant to the problem at hand. We searched through benchmarking catalogs such as PapersWithCode\footnote{\url{https://paperswithcode.com/datasets}}, competition websites such as Kaggle\footnote{\url{https://www.kaggle.com/datasets}}, academic search engines such as Google Scholar\footnote{\url{https://scholar.google.de}}, and  systematic review papers \cite{nasar2018information}. From these sources, we finally arrived at a representative list of 40 research datasets spanning the years 2011 to 2022.

%We elicit a use-case application of the ORKG-Dataset content type on research datasets of scientific IE. To create a collection of relevant research datasets to use as an example application, we manually searched benchmarking catalogs such as PapersWithCode\footnote{\url{https://paperswithcode.com/datasets}}, competition websites such as Kaggle\footnote{\url{https://www.kaggle.com/datasets}}, and one systematic review paper \cite{nasar2018information}. We thus arrived at a list of 40 research datasets spanning the years 2011 to 2022. Next, a team of four annotators specialized in the field of scientific information extraction followed an iterative methodology to identify the main contribution properties that align with the criteria of \textit{relevance}, \textit{usability}, and \textit{quality}. A structured representation for each research dataset was created with the properties described below. %that we recognized should be provided to dataset consumers to support the 3 main selection criteria of relevance, usability, and quality.

\subsection{Datasets Semantic Representation}

The next step was to create a structured representation for each of the research datasets based on their accompanying paper contributions. We, as a team of four annotators specialized in the field of scientific information extraction, followed an iterative methodology to identify the main contribution components. Our discussions resulted in the following main facets of information.

\begin{itemize}
    \item \textbf{Research Problems}: our initial search in scientific IE identified research datasets addressing various sub-problems. Some examples include citation classification, sentence classification, rhetorics annotation, relation extraction, coreference resolution, automated leaderboard construction, knowledge graph construction, scientific claim verification, text summarization, and text generation. This was modeled with the ORKG predicate \textit{research problem} (\url{https://orkg.org/property/P32}) and thus offers dataset consumers a clear indicator of \textit{relevance} to their specific tasks. 

    \item \textbf{Statistical attributes}: often when using machine learning methods, developers require additional statistics. E.g., for sentence classification datasets, how many sentences were annotated, and for how many documents. As such, we bundled nine relevant properties within a statistics template (\url{https://orkg.org/template/R220250}) to facilitate uniform modeling of this information across research datasets. Statistics information offers a direct \textit{usability} indicator to the dataset consumers. 

    \item \textbf{Quality}: one way to reflect a dataset's annotation quality is by inter-annotator agreement (IAA) scores, e.g. by using Cohen's kappa \cite{mchugh2012interrater}. This quality indicator can be specific to different information scopes, such as entities, relations, or sentences. To represent this information, we created the \textit{Data-centric result} template (\url{https://orkg.org/template/R220939}), which records the evaluation score and linked metric using the QUDT standardized methodology for evaluations (\url{https://qudt.org/schema/qudt/Quantity}). Additionally, the \textit{has evaluation item} property (\url{https://orkg.org/property/P71154}), modeled by the nested template \textit{Evaluation item} (\url{https://orkg.org/template/R221194}), allows specifying the granularity of annotations. This comprehensive unit of information serves as one direct indicator of dataset \textit{quality}. 

    \item \textbf{Performance Benchmarks}: Scholarly articles of research datasets report performance benchmarks. We identified this as an indirect \textit{quality} indicator for dataset consumers. We modeled this aspect with the help of the existing \textit{Leaderboard} template (\url{https://orkg.org/template/R107801}) which includes properties that allow specification of model name, model code URL, and allows specifying the score and metric per QUDT standards. 

    \item \textbf{Metadata}: technologies for efficient and effective reuse of ontological knowledge are one of the key success factors for developing ontology-based systems \cite{simperl2009reusing}. In this vein, we reused 19 relevant properties from the \url{https://schema.org/Dataset} content type. Some of the properties are \textit{name}, \textit{alternatename}, \textit{asseses}, \textit{description}, and \textit{URL}. Of particular interest is the URL property which is used to record the URL source where the dataset can be downloaded. These properties were modeled as the generic Dataset template (\url{https://orkg.org/template/R178304}) and could be uniformly applied across the 40 papers.
\end{itemize}

\begin{figure}
    \centering
    \includegraphics[width=\columnwidth]{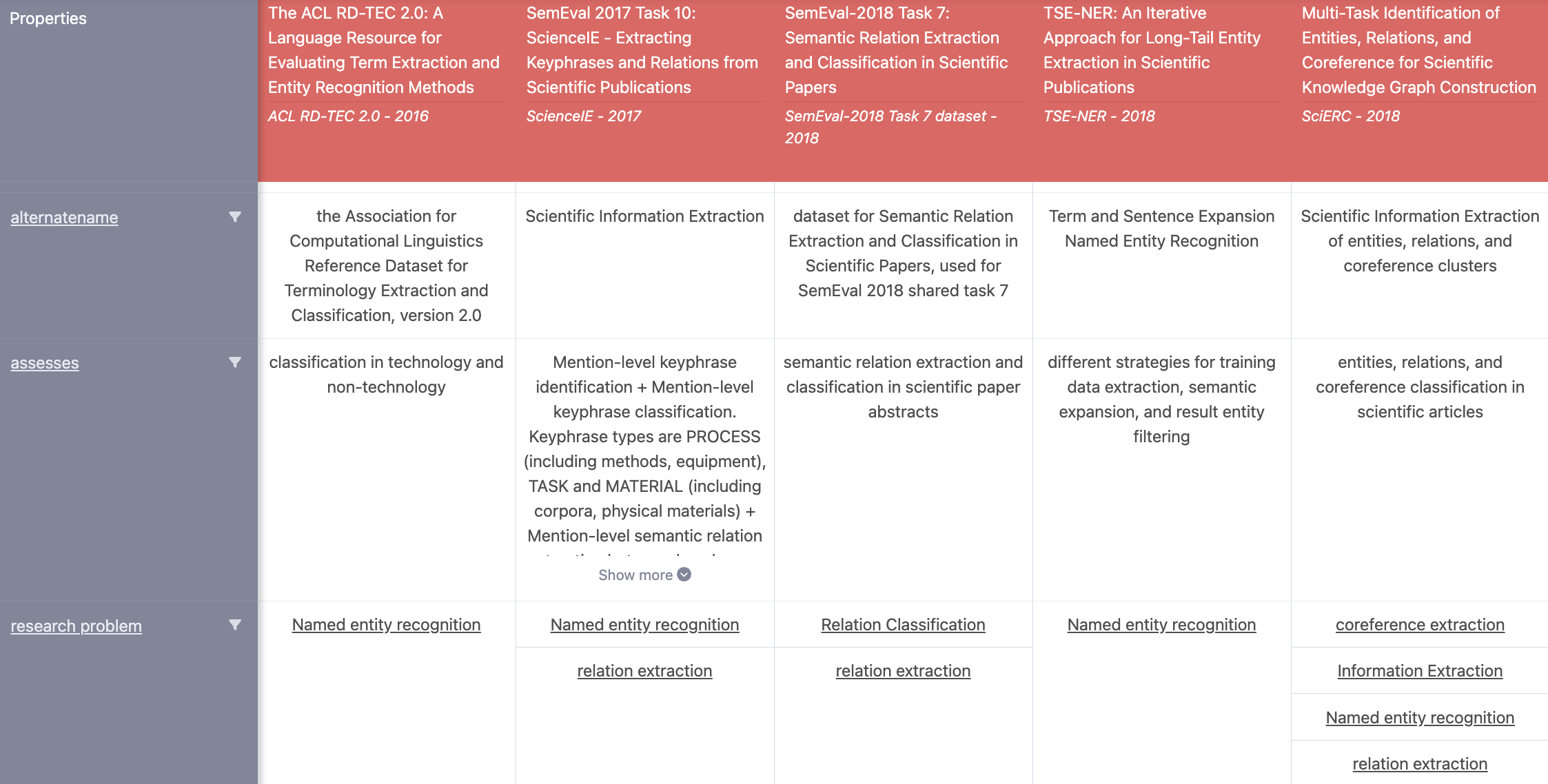}
    \caption{Excerpt of a screenshot of research datasets addressing scientific IE in the ORKG comparison view with structured metadata descriptions based on a set of properties defined as the ORKG-Dataset content type. The full comparison of 40 research datasets is accessible at \url{https://orkg.org/comparison/R280270/}.}
    \label{fig:screenshot}
\end{figure}

Putting all the pieces together, each of the 40 structured papers was finally typed with the ORKG-Dataset content type class \url{https://orkg.org/class/Dataset} which links to the generic Dataset metadata template. The result of our annotations is publicly accessible as an ORKG Comparison view \url{https://orkg.org/comparison/R280270/}. Figure~\ref{fig:screenshot} shows a partial screenshot.

\subsection{Customizable Querying}

A unique benefit of the semantic representation of research datasets in the ORKG is the ability to create customizable snapshots of the interconnected data graph in the context of the larger graph data capturing various other kinds of scholarly contributions. This presents users with advanced search and selection options that can be implemented via SPARQL queries. We elicit this in the following three scenarios.

\subsubsection{Bibliometric view.} With regard to bibliometrics, one could obtain detailed metadata about the authors, publication dates, and
citation numbers. For example, researchers could get citation statistics and obtain the most cited dataset for a particular task during a specific period, in order to get some idea of the design and specificities of datasets that had a high impact in the community.

\subsubsection{Dataset view.} Another example query is finding ground-truth datasets that could be used to train a model to solve a particular task. One such example query can be found in Figure \ref{query.1}. In addition, it is possible to filter for datasets that label particular entities, such
as “methods” or “materials” (Figure \ref{query.2}), and which match particular inter-annotator agreement scores if available. Most of the papers come with the actual URLs under which the published ground-truth datasets could be found. Thus, one could fetch the URL for all datasets of interest.

\begin{figure}[!tbh]
  \begin{Verbatim}[fontsize=\small]
SELECT ?task (GROUP_CONCAT(?dataset;separator=',') AS ?dataset) 
WHERE {
  res:R280270 pred:compareContribution ?contribution .
  ?contribution a class:Dataset ;
     rdfs:label ?dataset .
  ?contribution pred:P32/rdfs:label ?task 
}
GROUP BY ?task
  \end{Verbatim}
  %\vspace{-0.5cm}
  \caption{Example query to obtain a list of ground truth datasets and the tasks they address. Full query: \url{https://tinyurl.com/query-example-1}.}
  \label{query.1}
\end{figure}

\begin{figure}[!tbh]
  \begin{Verbatim}[fontsize=\small]
SELECT DISTINCT ?concept GROUP_CONCAT(?dataset;separator=',')
WHERE {
  res:R280270 pred:compareContribution ?contribution .
  ?contribution a class:Dataset ;
     rdfs:label ?dataset .
  ?contribution pred:P34062/rdfs:label ?concept .
  FILTER( ?concept = "Method"^^xsd:string 
    OR ?concept = "Research problem"^^xsd:string)
}
GROUP BY ?concept
  \end{Verbatim}
  \vspace{-0.5cm}
  \caption{Example query to filter for datasets that label ``Method'' and ``Research problem'' as labeled entity types in the ground truth. Full query: \url{https://tinyurl.com/query-example-2}.}
  \label{query.2}
\end{figure}

\subsubsection{SOTA view.} An integrated comparison table as ours also allows to uncover implicit relations between entities, for example, to search for competing, state-of-the-art machine learning models. Competing models are models that have been used to solve a similar NLP task, whereas a non-competing model would be a model that has not been used directly to solve the same task \cite{Daw2022ExtractionOC}. We could also filter out models that score above a particular evaluation metric, for example, models which score about 0.7 value in F1-score. Querying for the mostly used evaluation metrics could give researchers an idea of the metrics most widely used by the community.

\section{Conclusion}
\label{conclusion}

In this article, we presented the ORKG-Dataset content type as an approach for the semantic publishing of research datasets. The ORKG-Dataset content type breaks new ground in two main aspects: 1) in the studies of the transition from document-based to structured knowledge-based scholarly communication; and 2) of moving away from just metadata-based semantic descriptions of research datasets to including salient features of their accompanying scholarly publications as an enriched and more informative representation. Future developments of the ORKG-Dataset will apply it to more fields and thus make it more generalizable, as well as refine the properties to describe the quality of datasets using metrics beyond IAA. 

%\vspace{-0.5cm}
\section*{Funding Statement}
This work is funded by the Deutsche Forschungsgemeinschaft (DFG, German Research Foundation) – project number: NFDI4DataScience (460234259).

% FIXME: If accepted, a funding notice needs to be included somewhere that acknowledges the NFDI4DS funding (and perhaps other funding sources).

\bibliographystyle{vancouver.bst}
\bibliography{refs.bib}

\begin{thebibliography}{10}

\bibitem{cafarella2011structured}
Cafarella MJ, Halevy A, Madhavan J.
\newblock Structured data on the web.
\newblock Communications of the ACM. 2011;54(2):72-9.

\bibitem{manyika2014open}
Manyika J, Chui M, Farrell D, Van~Kuiken S, Groves P, Almasi~Doshi E. Open
  Data: Unlocking Innovation and Performance with Liquid Information| McKinsey
  \& Company; 2014.

\bibitem{verhulst2016open}
Verhulst S, Young A.
\newblock Open data impact when demand and supply meet key findings of the open
  data impact case studies.
\newblock Available at SSRN 3141474. 2016.

\bibitem{gregory2018understanding}
Gregory K, Cousijn H, Groth P, Scharnhorst AM, Wyatt SME.
\newblock Understanding Data Retrieval Practices: A Social Informatics
  Perspective.
\newblock Arxiv.org; 2018.

\bibitem{mayer2013big}
Mayer-Sch{\"o}nberger V, Cukier K.
\newblock Big data: A revolution that will transform how we live, work, and
  think.
\newblock Houghton Mifflin Harcourt; 2013.

\bibitem{brickley2019google}
Brickley D, Burgess M, Noy N.
\newblock Google Dataset Search: Building a search engine for datasets in an
  open Web ecosystem.
\newblock In: The World Wide Web Conference; 2019. p. 1365-75.

\bibitem{manghi2019openaire}
Manghi P, Bardi A, Atzori C, Baglioni M, Manola N, Schirrwagen J, et~al.
\newblock The OpenAIRE research graph data model.
\newblock Zenodo. 2019.

\bibitem{chapman2020dataset}
Chapman A, Simperl E, Koesten L, Konstantinidis G, Ib{\'a}{\~n}ez LD, Kacprzak
  E, et~al.
\newblock Dataset search: a survey.
\newblock The VLDB Journal. 2020;29(1):251-72.

\bibitem{koesten2020everything}
Koesten L, Simperl E, Blount T, Kacprzak E, Tennison J.
\newblock Everything you always wanted to know about a dataset: Studies in data
  summarisation.
\newblock International Journal of Human-Computer Studies. 2020;135:102367.

\bibitem{piwowar2010public}
Piwowar HA, Chapman WW.
\newblock Public sharing of research datasets: a pilot study of associations.
\newblock Journal of informetrics. 2010;4(2):148-56.

\bibitem{boukhelifa2017data}
Boukhelifa N, Perrin ME, Huron S, Eagan J.
\newblock How data workers cope with uncertainty: A task characterisation
  study.
\newblock In: Proceedings of the 2017 CHI Conference on Human Factors in
  Computing Systems; 2017. p. 3645-56.

\bibitem{gregory2019searching}
Gregory K, Groth P, Cousijn H, Scharnhorst A, Wyatt S.
\newblock Searching data: a review of observational data retrieval practices in
  selected disciplines.
\newblock Journal of the Association for Information Science and Technology.
  2019;70(5):419-32.

\bibitem{kern2015there}
Kern D, Mathiak B.
\newblock Are there any differences in data set retrieval compared to
  well-known literature retrieval?
\newblock In: Research and Advanced Technology for Digital Libraries: 19th
  International Conference on Theory and Practice of Digital Libraries, TPDL
  2015, Pozna{\'n}, Poland, September 14-18, 2015, Proceedings 19. Springer;
  2015. p. 197-208.

\bibitem{thoegersen2021researcher}
Thoegersen JL, Borlund P.
\newblock Researcher attitudes toward data sharing in public data repositories:
  a meta-evaluation of studies on researcher data sharing.
\newblock Journal of Documentation. 2021;78(7):1-17.

\bibitem{koesten2017trials}
Koesten LM, Kacprzak E, Tennison JF, Simperl E.
\newblock The Trials and Tribulations of Working with Structured Data: -a Study
  on Information Seeking Behaviour.
\newblock In: Proceedings of the 2017 CHI conference on human factors in
  computing systems; 2017. p. 1277-89.

\bibitem{wilkinson2016fair}
Wilkinson MD, Dumontier M, Aalbersberg IJ, Appleton G, Axton M, Baak A, et~al.
\newblock The FAIR Guiding Principles for scientific data management and
  stewardship.
\newblock Scientific data. 2016;3(1):1-9.

\bibitem{shotton2009semantic}
Shotton D.
\newblock Semantic publishing: the coming revolution in scientific journal
  publishing.
\newblock Learned Publishing. 2009;22(2):85-94.

\bibitem{auer2020improving}
Auer S, Oelen A, Haris M, Stocker M, D’Souza J, Farfar KE, et~al.
\newblock Improving access to scientific literature with knowledge graphs.
\newblock Bibliothek Forschung und Praxis. 2020;44(3):516-29.

\bibitem{oelen2020}
Oelen A, Jaradeh MY, Stocker M, Auer S.
\newblock Generate FAIR Literature Surveys with Scholarly Knowledge Graphs.
\newblock In: Proceedings of the ACM/IEEE Joint Conference on Digital Libraries
  in 2020. JCDL '20. New York, NY, USA: Association for Computing Machinery;
  2020. p. 97–106.
\newblock Available from: \url{https://doi.org/10.1145/3383583.3398520}.

\bibitem{gebru2021datasheets}
Gebru T, Morgenstern J, Vecchione B, Vaughan JW, Wallach H, Iii HD, et~al.
\newblock Datasheets for datasets.
\newblock Communications of the ACM. 2021;64(12):86-92.

\bibitem{jaradeh2019open}
Jaradeh MY, Oelen A, Farfar KE, Prinz M, D'Souza J, Kismih{\'o}k G, et~al.
\newblock Open research knowledge graph: next generation infrastructure for
  semantic scholarly knowledge.
\newblock In: Proceedings of the 10th International Conference on Knowledge
  Capture; 2019. p. 243-6.

\bibitem{haris2022persistent}
Haris M, Stocker M, Auer S.
\newblock Persistent Identification and Interlinking of FAIR Scholarly
  Knowledge.
\newblock arXiv preprint arXiv:220908789. 2022.

\bibitem{nasar2018information}
Nasar Z, Jaffry SW, Malik MK.
\newblock Information extraction from scientific articles: a survey.
\newblock Scientometrics. 2018;117(3):1931-90.

\bibitem{mchugh2012interrater}
McHugh ML.
\newblock Interrater reliability: the kappa statistic.
\newblock Biochemia medica. 2012;22(3):276-82.

\bibitem{simperl2009reusing}
Simperl E.
\newblock Reusing ontologies on the Semantic Web: A feasibility study.
\newblock Data \& Knowledge Engineering. 2009;68(10):905-25.

\bibitem{Daw2022ExtractionOC}
Daw S, Pudi V.
\newblock Extraction of Competing Models using Distant Supervision and Graph
  Ranking.
\newblock In: SDU@AAAI; 2022. .

\end{thebibliography}

\end{document}